\documentclass[onecollarge]{svjour2}       % onecolumn "king-size"
\smartqed  % flush right qed marks, e.g. at end of proof
\usepackage{psfrag,graphicx,epsfig,color, amsmath}
  \usepackage[percent]{overpic}

\journalname{Journal of Statistical Physics}
\begin{document}

\title{document}
\title{Advection diffusion equation with absorbing boundary}

\author{John Grant and Michael Wilkinson}

\institute{   Department of Mathematics and Statistics, \\
              The Open University,
              Walton Hall, \\
              Milton Keynes, MK7 6AA, \\
              England.\\
              \email{m.wilkinson@open.ac.uk}}

\date{Received: date / Accepted: date}

\maketitle

\begin{abstract}
We consider a spatially homogeneous advection-diffusion equation in which the diffusion 
tensor and drift velocity are time-independent, but otherwise general.  We derive asymptotic 
expressions, valid at large distances from a steady point source, for the flux onto a completely 
permeable boundary and onto an absorbing boundary. The absorbing case is treated by making 
a source of antiparticles at the boundary. In both cases there is an exponential decay as the 
distance from the source increases; we find that the exponent is the same for both boundary conditions. 

% 05.10.Gg	Stochastic analysis methods (Fokker-Planck, Langevin, etc.)
% 05.40.-a	Fluctuation phenomena, random processes, noise, and Brownian motion 
% 05.45.Df	Fractals 

\end{abstract}

\section{Introduction}
\label{sec: 1}

In this paper we discuss the solution of the advection-diffusion equation with a point source of intensity 
$\sigma(t)$ located at $\mbox{\boldmath$x$}_0$. In $d$ spatial dimensions the particle density  
$\rho(\mbox{\boldmath$x$},\mbox{\boldmath$x$}_0,t)$  at position 
$\mbox{\boldmath$x$}$ at time $t$ satisfies
\begin{equation}
\label{eq: 1.1}
\frac{\partial \rho}{\partial t}=-\sum_{i=1}^d v_i \frac{\partial \rho}{\partial x_i}+\sum_{i=1}^d
\sum_{j=1}^d D_{ij}\frac{\partial^2 \rho}{\partial x_i \partial x_j}+ \sigma(t) 
\delta (\mbox{\boldmath$x$}-\mbox{\boldmath$x$}_0)
\end{equation}
where the $v_i$ are components of the drift velocity $\mbox{\boldmath$v$}$,
the $D_{ij}$ are elements of the diffusion tensor ${\bf D}$ and both $\mbox{\boldmath$v$}$ and ${\bf D}$ 
are independent of position and time. 

Our objective is, for $d=2$, to determine the particle flux, $J_{\rm a}(x_2)$, from a steady source onto 
the boundary $x_1=0$ in the case where particles are absorbed on contact with the boundary
at $\mbox{\boldmath$x$}=(0,x_2)$. 
Dealing with an absorbing boundary is difficult when there is both advection and diffusion because 
there is no simple local boundary condition which  $\rho(\mbox{\boldmath$x$},\mbox{\boldmath$x$}_0,t)$ 
must satisfy. For this reason we also consider a reference problem where the boundary is 
completely permeable and has no effect on particles, the flux in this case is denoted by $J_{0}(x_2)$.

\begin{figure*}[!ht]
\begin{center}
  \begin{overpic}[width=0.75\textwidth, grid=false, tics=20]{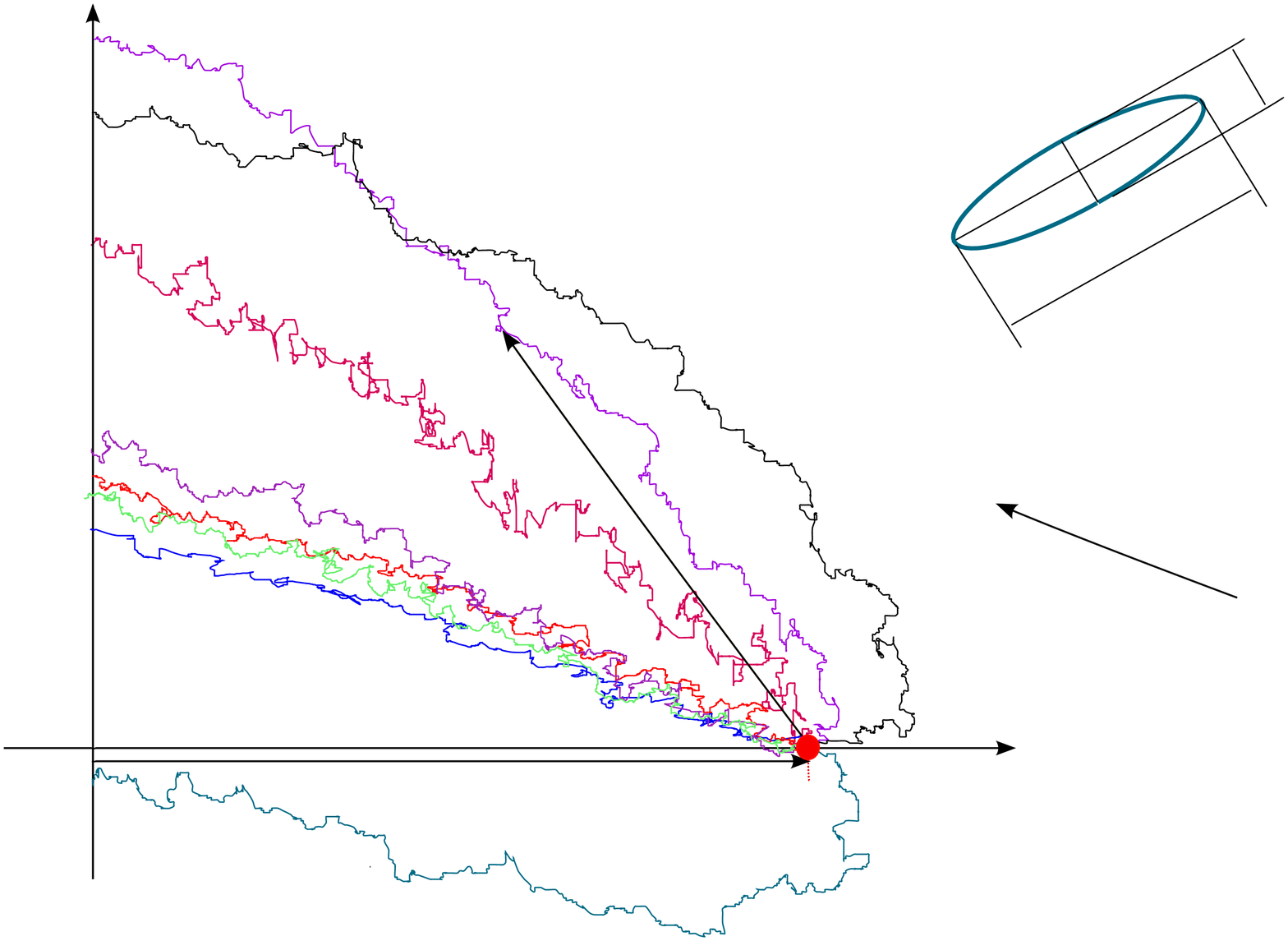} 
    \put(80,14){$x_1$}
    \put(5,74){$x_2$}
    \put(35,12){$x_0$}
    \put(86,31){$\mbox{\boldmath$v$}$}
    \put(80,65){$\bf D$}
		\put(86,49){$\sqrt{\mathcal{D}_1t}$}
		\put(98,67){$\sqrt{\mathcal{D}_2t}$}
		\put(41,40){$\mbox{\boldmath$\xi$}$}
  \end{overpic} 
\end{center}
\caption{
\label{fig: 1}
(Colour online). 
Particles created by a steady unit source at $(x_0,0)$ undergo advection with 
velocity $\mbox{\boldmath$v$}$ and diffusion with diffusion tensor ${\bf D}$.  
The displacement of a particle from the source is denoted by $\mbox{\boldmath$\xi$}$. 
The eigenvalues of ${\bf D}$ are $\mathcal{D}_1$  and $\mathcal{D}_2$ and the principal axes 
of  diffusion are the major and minor axes of the ellipse 
$\mbox{\boldmath$x$}^\textsc{T} \cdot {\bf D}\mbox{\boldmath$x$} = 1$.  In the absence 
of advection, after a time $t$ a  \lq droplet' comprising many particles initially concentrated 
at a point would occupy an ellipse with the orientation and dimensions shown in the figure.  
We wish to determine the flux density onto the absorbing boundary $x_1=0$ at $(0,x_2)$, for large $x_2$. 
}
\end{figure*}

The primary motivation for analysing this problem arose from studies of the 
structure of fractal measures generated by random dynamical systems which
can serve as models for particles in turbulent flows 
\cite{Fal+00,Bec+04,Pum+00}. 
A fractal measure can be characterised by the statistics of constellations of nearby 
points sampling the measure. The simplest case is where a constellation consisting
of just two points is characterised by the distance 
$\delta r$ between two trajectories. Defining $x_1=-{\rm ln}\,(\delta r/\xi)$, 
where $\xi$ is a correlation length of the flow, it is found that $x_1$ obeys an advection-diffusion 
equation when $x_1\gg 0$ \cite{Wil+10,Wil+12}. 
The point $x_1=0$ is both the location of an absorbing boundary 
and a source for trajectories entering the region $x_1>0$. In the region $x_1\gg 1$ the 
probability density of $x_1$  is an exponential function, which corresponds to a power-law 
distribution of $\delta r$. The exponent of this power-law determines the correlation dimension
of the fractal \cite{Wil+10,Wil+12}. 
If we consider a constellation formed by  
three trajectories, we can consider the shape statistics of triangles 
formed by triplets of points lying inside a small ball, of radius $\epsilon$ \cite{Wil+14}. 
The shape of a 
triangle may be characterised by the parameter $z={\cal A}/{\cal R}^2$, where ${\cal A}$ 
is its area and ${\cal R}$ its radius of gyration.  For a fractal measure generated by a 
random flow we showed that the variables $x_1=-{\rm ln}\ (\delta r/\xi)$ and $x_2=-{\rm ln}z$ evolve 
according to an advection-diffusion process in the region $x_1\gg 1$ and $x_2\gg 1$, with an 
absorbing boundary on the line $x_1=0$, and we are led to consider the problem 
illustrated in figure \ref{fig: 1}. This approach, described in \cite{Wil+14}, 
can be extended to show that other variables characterising shapes 
of constellations of nearby points can be described by a suitable advection-diffusion equation.

A second reason for studying this absorption problem is its relevance to the modelling 
of fallout plumes from events such as volcanic eruptions. The study of dispersion of 
dust and smoke in the atmosphere  has a long history (see \cite{Rob24,Sut32}). 
Ermak \cite{Erm77} appears to have been the first author to treat settling and 
deposition (that is drift and absorption) as well as diffusive dispersion. 
Stockie  \cite{Sto11} gives a recent review of this area. Our calculation would be relevant 
to fallout when the diffusive dispersion is anisotropic.

We investigate the flux onto the boundary $x_1=0$ at position $(0,x_2)$, 
from a steady point source at $(x_0,0)$. At large distance from the source the flux
onto an absorbing boundary has magnitude
\begin{equation}
\label{eq: 1.2}
J_{\rm a}(x_2) \sim A_{\rm a}(x_2)\exp[-\Phi_0(x_2)]
\end{equation}
where $\Phi_0(x_2)$ and $A_{\rm a}(x_2)$ both depend upon the source position $x_0$. 
We find that $\Phi_0(x_2)/x_2$ approaches finite limits as $x_2\to \pm \infty$ and that  
$A_{\rm a}(x_2)$ is asymptotic to a power law as $x_2\to \pm \infty$. For a transparent 
boundary the flux is of the same form, with $A_{\rm a}(x_2)$ replaced by another function 
$A_0(x_2)$ which decays more slowly as $|x_2|\to\infty$. Somewhat surprisingly, 
the exponent $\Phi_0(x_2)$ is the same for the transparent and absorbing boundary cases.

This paper is organised as follows. In section \ref{sec: 2}, we determine the density and 
flux for a permeable boundary, for general $\mbox{\boldmath$v$}$ and ${\bf D}$. 
The resulting expressions appear to be singular whenever ${\bf D}$ is a singular matrix, 
but in section \ref{sec: 2.3} we show that they give finite values even when ${\bf D}$ is a 
rank-one matrix. We consider the effect of an absorbing boundary in section \ref{sec: 3}, 
showing how an implicit expression for the flux can be written in terms of a source of 
\lq antiparticles'  which are created on the boundary. The one-dimensional case is solved 
explicitly, and we show that the method is in accord with results from the theory of first-passage 
time processes, discussed in \cite{Red01,Sch10}. In section \ref{sec: 4} the results from 
sections \ref{sec: 2} and  \ref{sec: 3} are combined to obtain the flux, in two dimensions, from 
a steady source at $\mbox{\boldmath$x$}_0= (x_0,0)$ onto an absorbing boundary along 
the $x_2$ axis.  A surprising cancellation ensures that the exponent $\Phi_{0}(x_2)$ is the 
same as that for the completely permeable boundary treated in section \ref{sec: 2}. We conclude 
in section \ref{sec: 5} with some numerical illustrations of the quality of the approximations 
and some comments on the structure of our asymptotic results.

\section{Particle density and flux for a permeable boundary}
\label{sec: 2}

In this section we consider the problem of calculating the steady state particle density and the 
corresponding flux arising from a steady point source. We then apply the results in two dimensions to
calculate the density and flux along a line, the $x_2$ axis. The line has no effect on the 
particles: it is a completely permeable or \lq transparent' boundary and the particles 
may cross it multiple times and at different locations. 

\subsection{Particle flux for a non-singular diffusion tensor}
\label{sec: 2.1}

In $d$  dimensions the free space propagator or Green's function for (\ref{eq: 1.1}) (that is, the
probability density for a particle released at $\mbox{\boldmath$x$}_0$ at time zero to reach 
$\mbox{\boldmath$x$}$ at time $t$)  is
\begin{eqnarray}
\label{eq: 2.1}
G_0 \left(\mbox{\boldmath$x$, \boldmath$x$}_0,t \right)
=  \left[
         \frac{\exp\left[-S(\mbox{\boldmath$\xi$},t)\right]}
				      { \sqrt{ {\rm det}({\bf D}) \left (4\pi t\right) ^{d} }} 
\right] \Theta\left(t\right)
\end{eqnarray}
where   $\Theta(t)$ is the Heaviside step function, $\mbox{\boldmath$\xi$}$ is the 
displacement vector $\mbox{\boldmath$x$}-\mbox{\boldmath$x$}_0$ and
\begin{equation}
\label{eq: 2.2}
S(\mbox{\boldmath$\xi$},t)= \frac{(\mbox{\boldmath$\xi$}-\mbox{\boldmath$v$}t)
\cdot {\bf D}^{-1}(\mbox{\boldmath$\xi$}-\mbox{\boldmath$v$}t)}{4t}
\ .
\end{equation}
The particle density at point $\mbox{\boldmath$x$} $ at time $t$ is therefore
\begin{equation}
\label{eq: 2.3}
\rho_0 \left(\mbox{\boldmath$x$,\boldmath$x$}_0, t \right) =
\int_{-\infty}^{t} {\rm d}t^\prime \ \sigma(t^\prime) G_0(\mbox{\boldmath$x$,\boldmath$x$}_0,t-t^\prime) 
\ . 
\end{equation}
The flux vector $\mbox{\boldmath$J$}_0 ( \mbox{\boldmath$x$}, \mbox{\boldmath$x$}_0,t)$ is 
\begin{equation}
\label{eq: 2.5}
\mbox{\boldmath$J$}_0 (\mbox{\boldmath$x$},\mbox{\boldmath$x$}_0,t) =
\rho_{0}(\mbox{\boldmath$x$},\mbox{\boldmath$x$}_0,t) \ \mbox{\boldmath$v$} 
- \mbox{\boldmath$D$} \  \mbox{\boldmath$\nabla$}\rho_{0}
(\mbox{\boldmath$x$},\mbox{\boldmath$x$}_0,t) 
\
\end{equation}
where the gradient is with respect to $\mbox{\boldmath$x$}$. 
Now consider the steady-state density $\rho_0(\mbox{\boldmath$x$},\mbox{\boldmath$x$}_0)$ 
and flux $\mbox{\boldmath$J$}_0(\mbox{\boldmath$x$},\mbox{\boldmath$x$}_0)$ due to a 
steady source $\sigma(t)=\sigma_0$. Using (\ref{eq: 2.3}) and employing (\ref{eq: 2.1}), (\ref{eq: 2.2})
to evaluate the gradient,
\begin{equation}
\label{eq: 2.7}
\mbox{\boldmath$D$} \mbox{\boldmath$\nabla$}\rho_0 (\mbox{\boldmath$x$},\mbox{\boldmath$x$}_0,t) 
= \sigma_{0} \int_{-\infty}^{0} {\rm d}t' \ \mbox{\boldmath$D$} 
\mbox{\boldmath$\nabla$}
G_0(\mbox{\boldmath$x$,\boldmath$x$}_0,t-t')  
= - \sigma_{0} \int_{-\infty}^{0} {\rm d}t' \   G_0(\mbox{\boldmath$x$,\boldmath$x$}_0,t-t') 
\frac{1}{2(t-t^\prime)} \left( \mbox{\boldmath$\xi$} - \mbox{\boldmath$v$}(t- t^\prime)   \right)
\end{equation}
giving
\begin{equation}
\label{eq: 2.8}
\mbox{\boldmath$J$}_0 (\mbox{\boldmath$x$},\mbox{\boldmath$x$}_0) = 
\sigma_{0} \int_{-\infty}^{0} {\rm d}t \   G_0(\mbox{\boldmath$x$,\boldmath$x$}_0,t) 
\mbox{\boldmath$v$}_{\rm eff} (\mbox{\boldmath$x$},\mbox{\boldmath$x$}_0,t) 
\end{equation}
where the effective velocity is
\begin{equation}
\label{eq: 2.9}
 \mbox{\boldmath$v$}_{\rm eff} (\mbox{\boldmath$x$},\mbox{\boldmath$x$}_0,t) \equiv 
\frac{1}{2} \left( \frac{\mbox{\boldmath$x$} - \mbox{\boldmath$x$}_0}{t} + \mbox{\boldmath$v$}\right)
\ .
\end{equation}

\subsection{Asymptotic values of particle density and flux}
\label{sec: 2.2}

It is not possible to express the integrals in equations (\ref{eq: 2.3}) and 
(\ref{eq: 2.8}) exactly in a closed form. However, for large 
$|\mbox{\boldmath$\xi$}|=|\mbox{\boldmath$x$}-\mbox{\boldmath$x$}_0|$, these integrals 
are dominated by contributions from a neighbourhood of the critical point $t^\ast$ at which 
$S(\mbox{\boldmath$\xi$},t)$ has a minimum. The method of Laplace can be applied 
to estimate these integrals (we comment later on the identity of the large parameter 
of the theory). The Laplace method gives the following estimates for the steady-state 
density and flux with a permeable boundary
\begin{eqnarray}
\label{eq: 2.11}
\rho_0(\mbox{\boldmath$x$, \boldmath$x$}_0)&\sim &\gamma  \sigma_0   
G_0 \left( \mbox{\boldmath$x$,\boldmath$x$}_0,t^\ast \right) 
\nonumber \\
\mbox{\boldmath$J$}_0 (\mbox{\boldmath$x$},\mbox{\boldmath$x$}_0) &\sim &   
\rho_0(\mbox{\boldmath$x$, \boldmath$x$}_0) \mbox{\boldmath$v$}_{\rm eff} 
(\mbox{\boldmath$x$},\mbox{\boldmath$x$}_0,t^\ast )
\ .
\end{eqnarray}
where $\gamma$ is the Gaussian integral 
\begin{equation}
\label{eq: 2.13}
\gamma = \int_{-\infty}^\infty {\rm d} u \ \exp \left[-\frac{1}{2} 
\frac{\partial^2 S(\mbox{\boldmath$\xi$},t^\ast)}{\partial t^2} u^2 \right] \ .
\end{equation}
The condition that  $S$  be stationary with respect to $t$ is 
\begin{equation}
\label{eq: 2.14}
\frac{\partial S}{\partial t}(\mbox{\boldmath$\xi$},t^\ast) =
\frac{\mbox{\boldmath$v$}\cdot{\bf D}^{-1}\mbox{\boldmath$v$} (t^\ast)^2 -
\mbox{\boldmath$\xi$}\cdot{\bf D}^{-1}\mbox{\boldmath$\xi$}}{4 (t^\ast)^2} = 0
\end{equation}
giving 
\begin{equation}
\label{eq: 2.15}
t^\ast = \sqrt{\frac{\mbox{\boldmath$\xi$}\cdot{\bf D}^{-1}
\mbox{\boldmath$\xi$}}{\mbox{\boldmath$v$}\cdot{\bf D}^{-1}\mbox{\boldmath$v$}}}  
\ .
\end{equation}
Since  ${\bf D}$ is symmetric 
\begin{equation}
\label{eq: 2.16}
S(\mbox{\boldmath$\xi$},t^\ast) = \frac{1}{2} \left( \sqrt{\mbox{\boldmath$\xi$}\cdot{\bf D}
^{-1}\mbox{\boldmath$\xi$}} \sqrt{\mbox{\boldmath$v$}\cdot{\bf D}^{-1}\mbox{\boldmath$v$}} 
-\mbox{\boldmath$\xi$}\cdot{\bf D}^{-1}\mbox{\boldmath$v$} \right) 
\end{equation}
Also, the second derivative and the Gaussian integral (\ref{eq: 2.13}) are
\begin{equation}
\label{eq: 2.18}
\frac{\partial^2 S}{\partial t^2}(\mbox{\boldmath$\xi$},t^\ast) =\frac{1}{2}\mbox{\boldmath$v$}\cdot
{\bf D}^{-1}\mbox{\boldmath$v$} \sqrt{\frac{\mbox{\boldmath$v$}\cdot{\bf D}^{-1}\mbox{\boldmath$v$}}
{\mbox{\boldmath$\xi$}\cdot{\bf D}^{-1}\mbox{\boldmath$\xi$}}}
\ ,\ \ \ 
\gamma = \frac{2 \sqrt{\pi}\left( \mbox{\boldmath$\xi$}\cdot{\bf D}^{-1}
\mbox{\boldmath$\xi$} \right)^\frac{1}{4}} {\left( \mbox{\boldmath$v$}\cdot{\bf D}^{-1}
\mbox{\boldmath$v$}  \right)^\frac{3}{4}}
\ .
\end{equation}
Substituting these results into (\ref{eq: 2.11}) gives
\begin{equation}
\label{eq: 2.19}
\rho_0(\mbox{\boldmath$x$,\boldmath$x$}_0)  \sim  \sigma_0 
A_0(\mbox{\boldmath$x$,\boldmath$x$}_0) 
\exp \left[ -\Phi_0 \left( \mbox{\boldmath$x$,\boldmath$x$}_0 \right)  \right]
\end{equation}
where 
\begin{equation}
\label{eq: 2.20}
\Phi_0(\mbox{\boldmath$x$,\boldmath$x$}_0)= S(\mbox{\boldmath$\xi$},t^\ast)
\ ,\ \ \ \ 
A_0(\mbox{\boldmath$x$,\boldmath$x$}_0) =  
\frac{ \left( \mbox{\boldmath$\xi$}\cdot{\bf D}^{-1}\mbox{\boldmath$\xi$} \right)^{\left(\frac{1-d}{4} \right)}
\left( \mbox{\boldmath$v$}\cdot{\bf D}^{-1}\mbox{\boldmath$v$} \right)^{\left(\frac{d-3}{4}\right)} }
{\sqrt{ {\rm det} \left( {\bf D} \right) \left( 4 \pi \right)^{(d-1)}}}
\ .
\end{equation}
From equation (\ref{eq: 2.11}) when $|\mbox{\boldmath$\xi$}|$ is large the asymptotic value of the 
steady state flux can be written as
\begin{equation}
\label{eq: 2.22}
\mbox{\boldmath$J$}_0 (\mbox{\boldmath$x$},\mbox{\boldmath$x$}_0) \sim 
\frac{\sigma_0}{2}  A_0(\mbox{\boldmath$x$,\boldmath$x$}_0) \left[ \mbox{\boldmath$v$} +
\left( \sqrt{\frac{\mbox{\boldmath$v$}\cdot{\bf D}^{-1}\mbox{\boldmath$v$}} {\mbox{\boldmath$\xi$}\cdot
{\bf D}^{-1}\mbox{\boldmath$\xi$}} } \right) \mbox{\boldmath$\xi$}  
\right]
\exp \left[ -\Phi_{0}(\mbox{\boldmath$x$},\mbox{\boldmath$x$}_0)  \right] 
\ .
\end{equation}

On the streamline {\sl directly} downwind of the source the displacement vector 
$\mbox{\boldmath$\xi$}=\mbox{\boldmath$x$}-\mbox{\boldmath$x$}_0 = \lambda \mbox{\boldmath$v$}$ 
for some $ \lambda>0 $ so that \ $\Phi_0(\mbox{\boldmath$x$,\boldmath$x$}_0)= 0$, 
$P_0(\mbox{\boldmath$x$,\boldmath$x$}_0) \sim  \sigma_0 A_0(\mbox{\boldmath$x$,\boldmath$x$}_0)$ 
and $\mbox{\boldmath$J$}_0 (\mbox{\boldmath$x$},\mbox{\boldmath$x$}_0) 
\sim  \sigma_0 A_{0}(\mbox{\boldmath$x$},\mbox{\boldmath$x$}_0)  \mbox{\boldmath$v$}$.  
Therefore, directly downwind of the source point, there is no exponential reduction of the steady state 
particle density or the particle flux and the flux is directed along the drift vector. However, there is an 
algebraic reduction in both as the distance  $|\mbox{\boldmath$\xi$}|$ from the source increases.  
On any other ray starting from  $\mbox{\boldmath$x$}_0$ then as  $|\mbox{\boldmath$\xi$}|$ 
increases $\Phi_0$ increases and the particle density and flux undergo the same exponential 
reduction along the ray. The algebraic coefficients in the particle density and the flux are 
asymptotic to the same power law: $|\mbox{\boldmath$\xi$}|^{\frac{1-d}{2}}$. 

The Laplace method provides an estimate of integrals of the form
\begin{equation}
\label{eq: 2.23}
I(\lambda)=\int_a^b {\rm d}\tau f(\tau)\exp[-\lambda S(\tau)]
\end{equation}
in the limit as $\lambda\to \infty$, which assumes that $S(\tau)$ has a 
minimum in the interval $[a,b]$ at $\tau^\ast$. It is not immediately 
clear that (\ref{eq: 2.3}) is in this form. To see the connection define
$\tau=t/t^\ast$, where $t^\ast$ was given by (\ref{eq: 2.15}). Equation 
(\ref{eq: 2.3}) can then be expressed in a form similar to (\ref{eq: 2.23}),
where the large parameter is
\begin{equation}
\label{eq: 2.24}
\lambda=\sqrt{\mbox{\boldmath$\xi$}\cdot {\bf D}^{-1}\mbox{\boldmath$\xi$}}
\sqrt{\mbox{\boldmath$v$}\cdot {\bf D}^{-1}\mbox{\boldmath$v$}}
\ .
\end{equation}

In the two-dimensional case, which is our primary concern, the magnitude of the flux onto 
the boundary $x_1=0$ is the magnitude of first component of $\mbox{\boldmath$J$}_0$. At position $x_2$, this is
\begin{equation}
\label{eq: 2.25}
J_0(x_2)\equiv |[\mbox{\boldmath$J$}_0(x_2)]_1| = A_0(x_2)\exp[-\Phi_0(x_2)]  
\end{equation}
where  $\Phi_{0}(x_2) \equiv  \Phi_{0}((0,x_2),(x_0,0))$, and where 
\begin{equation}
\label{eq: 2.27}
A_0(x_2) \sim \frac{1}{\sqrt{|x_2|}}
\  .
\end{equation}

\subsection{Rank-one diffusion tensor}
\label{sec: 2.3}

The expressions for the density and flux which we have obtained contain ${\bf D}^{-1}$, 
the inverse of the diffusion tensor. We might therefore expect the analysis in the preceding sections to be valid only when ${\bf D}$ is nonsingular. However we now show that 
the results continue to hold for the case where ${\bf D}$ is a rank one matrix.

In general since ${\bf D}$ is symmetric it may be diagonalised by a suitable rotation of the axes.
Because we are, at this point, still considering a permeable boundary we can apply a 
rotation to diagonalise ${\bf D}$ without loss of generality. We will  therefore consider the 
case where  ${\bf D}$ is of the form
\begin{equation}
\label{eq: 2.32}
{\bf D}=
\begin{bmatrix}
D_{11} & 0 \\ 0 & \epsilon
\end{bmatrix}
\end{equation}
and take the limit as $\epsilon \to 0$.  This limiting case corresponds to diffusive motion 
with drift in the $x_1$ direction, with a diffusion constant $D_{11}$ and drift velocity $v_1$, 
together with drift with velocity $v_2$, and almost no diffusion, in the $x_2$ direction.

With  ${\bf D}$ given by (\ref{eq: 2.32})  and $\epsilon$ small but non-zero, then using 
equation (\ref{eq: 2.19}) to calculate the particle density far from the source 
we find that the terms in $\epsilon^{-1}$ cancel.  
Neglecting terms of order $\epsilon^2$  we find that the result is independent 
of $\epsilon$. At $(0,x_2)$ the density is
\begin{equation}
\label{eq: 2.33}
\rho_0 (x_2)  \sim 
\frac{ \sigma_{0} }{2 \sqrt{\pi D_{11} x_{2} v_{2}}} 
\exp  \left[
            - \frac{ x_{2} v_{1}^2}{4 D_{11} v_{2}} 
			\right] 
\ .
\end{equation}
We now show that this is in fact equal to the probability density for the case 
where $\epsilon=0$. If $\epsilon=0$ the displacement from $x_0$ in the $x_1$ direction 
is due to one-dimensional diffusion with drift, with diffusion coefficient $D_{11}$ and drift 
velocity $v_1$. The propagator for the $x_1$ component of the motion is therefore
\begin{equation}
\label{eq: 2.34}
G_{0} \left(x_1,x_0,t \right) = \frac{1}{\sqrt{4 \pi D_{11} t}} 
\exp \left[ 
            - \frac {\left( x_1 -x_0 -v_1t \right)^2}{4D_{11}t} 
		\right] \Theta(t)
\ .
\end{equation}
Further, when $\epsilon = 0$ there is no diffusion in the $x_2$ direction and, at time $t$, 
the $x_2$ coordinate of the particle is given by $x_2=v_2 \ t$.  The propagator for the two 
dimensional motion when $\epsilon = 0$ is therefore
\begin{equation}
\label{eq: 2.35}
G \left( \left(x_1,x_2 \right), \left(x_0,0 \right), t \right) = 
G_{0} \left(x_1,x_0,t \right) \delta \left(x_2 - v_2 t \right)
\ .
\end{equation}
From (\ref{eq: 2.3}), for a source of intensity $\sigma(t) = \sigma_0 \Theta(t)$  the particle density 
at $\mbox{\boldmath$x$}=\left( x_1,x_2 \right)$ from a steady source at $\mbox{\boldmath$x$}_0$ 
is 
\begin{eqnarray}
\label{eq: 2.36}
\rho_0(\mbox{\boldmath$x$},\mbox{\boldmath$x$}_0)& =& 
\frac{\sigma_0}{\sqrt{4 \pi D_{11}}}  \int_{0}^\infty  \frac{{\rm d} t'}{\sqrt{t'}} \ 
\exp { \left[ 
               - \frac{\left( x_{1} - x_{0} - v_{1} t^\prime \right)^2 }{4 D_{11}  t^\prime} 
			 \right] } \delta \left( x_2 -v_2 t^\prime \right)
\nonumber \\
& =& 
 \frac{\sigma_0}{\sqrt{4 \pi D_{11} x_2 v_2 } }
\exp { \left[ 
              - \frac{ v_2\left(x_1 - x_0 -v_{1}x_{2}/v_{2} \right)^2}{4 D_{11} x_2 } 
			 \right] } 
\end{eqnarray}
and on the boundary far from the source, since $x_1=0$ and $x_2 \gg x_0$, this reduces to (\ref{eq: 2.33}).

\section{Antiparticle method for an absorbing boundary}
\label{sec: 3}

\subsection{Integral equation for the absorption flux}
\label{sec: 3.1}

It is straightforward to perform a Monte Carlo simulation of the advection diffusion process 
with an absorbing boundary: the particles are simply removed from the simulation 
whenever they cross the boundary.  In contrast when solving a Fokker-Planck equation, 
such as equation (\ref{eq: 1.1}) for a particle density, it is not possible to write down a 
local boundary condition which implies that particles colliding with a boundary are absorbed.  
However, we can add an additional source term to the Fokker-Planck equation and, in principle, 
this allows us to describe an absorbing boundary by adding a suitable source of \lq antiparticles' 
at the boundary. After emission at the boundary the antiparticles propagate in the same way as 
particles do but, to model absorption, the antiparticle density is subtracted from the particle density. 
Since this approach does not rely on any geometrical symmetries of the system it is  
more general than approaches which use the method of images.

In the two-dimensional case suppose that position on the boundary $\Sigma$ can 
be  parameterised by  the arc length $s$ and let $j(\mbox{\boldmath$x$}', t')$ 
be the rate of antiparticle creation {\sl per unit length} at point $\mbox{\boldmath$x$}'$ 
and time $t'$. Then the element, ${\rm d}s$ of the boundary at point $\mbox{\boldmath$x$}'$ is 
a source of antiparticles with intensity $j(\mbox{\boldmath$x$}', t') {\rm d}s$  and the 
antiparticles created at the boundary contribute an amount 
\begin{equation}
\label{eq: 3.1}
-\int_{\Sigma}  {\rm d} s \ \int^t_{-\infty} {\rm d}t' \ j(\mbox{\boldmath$x$}',t')
G_0( \mbox{\boldmath$x$},\mbox{\boldmath$x$}' ,t-t') 
\end{equation}
to the overall density at point $\mbox{\boldmath$x$}$ at time $t$.

Each antiparticle is created in response to the {\sl first arrival} of a particle at the 
boundary. The intensity of the antiparticle source is therefore 
\begin{equation}
\label{eq: 3.2}
j(\mbox{\boldmath$x$}', t') 
= \mbox{\boldmath$n$} \left( \mbox{\boldmath$x$}'\right) \cdot 
\mbox{\boldmath$J$}_{\rm a}( \mbox{\boldmath$x$}', \mbox{\boldmath$x$}_0,t') 
\end{equation}
where ${\mbox{\boldmath$J$}_{\rm a} ( \mbox{\boldmath$x$}', \mbox{\boldmath$x$}_0,t'})$ is the 
absorption flux  at $\mbox{\boldmath$x$}'$ and $t'$ due to a particle source at 
$\mbox{\boldmath$x$}_0$ and ${\bf n}( \mbox{\boldmath$x$})$ is the unit 
normal on the boundary at $\mbox{\boldmath$x$}^\prime$, pointing out of the region of 
diffusion-advection. The particle density in the presence of an absorbing boundary 
$\rho_a(\mbox{\boldmath$x$},\mbox{\boldmath$x$}_0,t)$ is therefore
\begin{equation}
\label{eq: 3.3}
\rho_{\rm a}(\mbox{\boldmath$x$},\mbox{\boldmath$x$}_0,t) =  \rho_0(\mbox{\boldmath$x$},
\mbox{\boldmath$x$}_0,t) -  \int_{\Sigma} {\rm d} s\  \int_{-\infty}^t {\rm d}t' \ {\bf n}
( \mbox{\boldmath$x$}')  \cdot \mbox{\boldmath$J$}_a (\mbox{\boldmath$x$}', 
\mbox{\boldmath$x$}_0,t') \ G_0(\mbox{\boldmath$x$},\mbox{\boldmath$x$}',t-t') 
\ .
\end{equation}

\subsection{Exact absorption flux in one dimension}
\label{sec: 3.2}

We illustrate the solution of the integral equation (\ref{eq: 3.3}) in one dimension and derive  
an exact expression for the flux density $\mbox{\boldmath$J$}_{\rm a}(0,x_0,t)$ of 
particles absorbed at the origin, $x=0$, arising from a source of strength $\sigma(t) = \delta(t)$ 
localised at $x_0$, with $x_0 > 0$. 

If  $D$ is the diffusion coefficient and $\mbox{\boldmath$v$} = -v \mbox{\boldmath$i$} $ is 
the drift velocity (so that the particles drift towards the left) then for $t>0$ the free 
propagator, representing propagation without absorption, is
\begin{equation}
\label{eq: 3.9}
G_0(x,x_0,t)=\frac{1}{\sqrt{4\pi Dt}}\exp\left[-\frac{(x-x_0+vt)^2}{4Dt}\right]\Theta(t)
\end{equation}
and the corresponding free flux density  is $\mbox{\boldmath$J$}_0(x,x_0,t) \equiv -J_0(x,x_0,t)
\mbox{\boldmath$i$} $ where the flux \textsl{onto} the boundary  (i.e. to the left) is
\begin{eqnarray}
\label{eq: 3.10}
J_0(x,x_0,t)&=&vG_0(x,x_0,t) + D\frac{\partial G_0}{\partial x}(x,x_0,t)
\nonumber \\
&=& \frac{ 1 }{\sqrt{4\pi Dt}}\left( \frac{x_0-x+vt}{2t} \right)\exp\left[-\frac{(x-x_0+vt)^2}{4Dt}\right] 
\ .
\end{eqnarray}
The one-dimensional form of (\ref{eq: 3.3}), giving the particle density at point $x$ at time $t$ 
for a system with an absorbing boundary at $x=0$, is
\begin{equation}
\label{eq: 3.11}
\rho_{\rm a}(x,x_0,t)=\rho_0(x,x_0,t)-\int_0^t {\rm d}t'\ J_{\rm a}(0, x_0, t') G_0(x,0,t-t')
\ .
\end{equation}
The corresponding flux is
\begin{equation}
\label{eq: 3.12}
J_{\rm a}(x,x_0,t)=  v\rho_{\rm a}(x,x_0,t) + D\frac{\partial \rho_{\rm a}}{\partial x}(x,x_0,t)
\end{equation}
and using (\ref{eq: 3.11}) in (\ref{eq: 3.12}) gives
\begin{equation}
\label{eq: 3.13}
J_{\rm a}(x,x_0,t)=J_0(x,x_0,t)  -\int_0^t{\rm d}t'\ J_{\rm a}(0,x_0,t')\,J_0(x,0,t-t')  
\ .
\end{equation}
This equation expresses the absorption flux at time $t$ from an ejection of particles 
at $t=0$ as the sum of a direct flux from the source and a term resulting from the 
creation of \lq antiparticles' due to the flux which reached the boundary at the earlier 
time $t'$.  
Since $J_0(x,x_0,t)$ is known (from (\ref{eq: 3.10})), equation (\ref{eq: 3.13}) is a linear 
Volterra integral equation of the second kind for $J_{\rm a}(x,x_0,t)$, with kernel 
$J_0(x,0,t)$. It may be solved for $J_a$ using the method of Laplace transforms. 
Noting that the integral in the equation is a convolution, the Laplace transform of 
(\ref{eq: 3.13}), in the time variable, is 
\begin{equation}
\label{eq: 3.14}
\bar J_{\rm a}(x,x_0,s) = \bar J_0(x,x_0,s)- \bar J_{\rm a}(0,x_0,s) \bar J_0(x,0,s)
\end{equation}
where $\bar J_{\rm a}$ and $\bar J_0$ are the Laplace transforms of $J_{\rm a}$ 
and $J_0$, respectively. 

Since $x=0$ is an absorbing boundary we must have $J_a(x,x_0,t) = 0$ for $x<0$, 
therefore, to determine  $J_{\rm a}(0,x_0,t)$, we consider
\begin{equation}
\label{eq: 3.15}
\bar J_{\rm a}(\epsilon,x_0,s) = \bar J_0(\epsilon,x_0,s)- \bar J_{\rm a}(0,x_0,s) 
\bar J_0(\epsilon,0,s)
\end{equation}
in the limit as ${\epsilon \to 0^+}$. The Laplace transform of $J_0(x,x_0,t)$ is
\begin{equation}
\label{eq: 3.16}
\bar J_0(x,x_0,s)= 
\left( \frac{1  \mp \alpha}{2\alpha}\right) \exp{ \left[ -\left( \frac{\xi v}{2D} \right) 
\left(1 \pm \alpha \right) \right]}
\end{equation}
according as $\xi=x-x_0$ is  $\pm $ve and where
\begin{equation}
\label{eq: 3.17}
\alpha  \equiv \sqrt{1 + \frac{4Ds}{v^2}}
\ .
\end{equation}
So equation (\ref{eq: 3.15}) becomes 
\begin{equation}
\label{eq: 3.18}
\bar J_{\rm a}(\epsilon,x_0,s) + \bar J_{\rm a}(0,x_0,s)  \left( \frac{1-\alpha}{2\alpha} \right) 
\exp \left[ - \frac{\epsilon v}{2D}(1+\alpha) \right] =  \left( \frac{1+\alpha}{2\alpha} \right) \exp 
\left[ - \frac{(\epsilon -x_0)v}{2D}(1 - \alpha) \right]
\end{equation}
and in the limit as $\epsilon \to 0^+$ this gives
\begin{equation}
\label{eq: 3.19}
\bar J_{\rm a}(0,x_0,s) =  \exp \left[ - \frac{x_0 v}{2D} \left( \alpha-1 \right) \right] =
\exp \left[ - \frac{x_0v}{2D} \left( \sqrt{1+\frac{4Ds}{v^2}}-1 \right) \right]
\ .
\end{equation}
Alternatively, we can consider the limit as $\epsilon \to 0^-$:
in this case there is no absorption flux at $x=0^-$ because particles reaching $x=0$ have been absorbed, and (\ref{eq: 3.15}) is replaced by
\begin{equation}
\label{eq: 3.15a}
0= \bar J_0(\epsilon,x_0,s)- \bar J_{\rm a}(0,x_0,s) 
\bar J_0(\epsilon,0,s)
\ .
\end{equation}
It is readily verified that using (\ref{eq: 3.15a}) instead of (\ref{eq: 3.15}) leads to the same
result.  

Inverting the Laplace transform gives the exact expression for the rate of absorption onto the boundary:
\begin{equation}
\label{eq: 3.20}
J_{\rm a}(0,x_0,t)=\frac{1}{\sqrt{4\pi D}}\frac{x_0}{t^{3/2}}\exp\left[-\frac{(x_0-vt)^2}{4Dt}\right]
\ .
\end{equation}
Distributions of this form are in a class which are sometimes referred inverse Gaussian or Wald 
distributions.

\section{Flux onto an absorbing boundary in two dimensions}
\label{sec: 4}

\subsection{ The method of calculation}
\label{sec:4.1}

We now apply the results derived in sections (\ref{sec: 2.3}) and (\ref{sec: 3.2}) to the case of 
absorption on a line in two dimensions.  Absorption at the point $( 0,x_2)$ requires that the 
particle  {\sl first} arrives at the boundary $x_1=0$ at ordinate $x_2$. We will consider 
the problem in terms of the first arrival time at the boundary and the distribution of the 
absorption ordinate conditional on that arrival time. 

In the following we introduce several probability density functions. We shall adopt 
the convention that the probability density function (PDF) of a random variable $x$ is
denoted by a function $P_x$, so that the probability that $x$ lies in the 
interval $[x,x+{\rm d}x]$ is $P_x(x)\,{\rm d}x$.
The first arrival time is determined by the $x_1$ component of the motion alone. This is a 
one-dimensional advection-diffusion process, with diffusion constant $D_{11}$, drift velocity 
$v_1=-v$ and starting from initial point $x_0$. The PDF for the 
first arrival time is the flux obtained above. In this context equation (\ref{eq: 3.20}) becomes 
\begin{equation}
\label{eq: 4.1}
P_{t}(t)=\frac{x_0}{\sqrt{4\pi D_{11}t^3}}\exp \left[- \frac{(x_0-v_1 t)^2}{4D_{11}t}\right]
\ .
\end{equation}
The absorption ordinate is a random variable, $x_2$, whose PDF may be obtained
from $P_{x_2|t}(x_2,t)$, the probability density for $x_2$ conditional upon the first arrival time 
$t$: 
\begin{equation}
\label{eq: 4.2}
P_{x_2}(x_2)=\int_0^\infty {\rm d}t\ P_{x_2|t}(x_2,t) \ P_{t}(t)
\ .
\end{equation}
The conditional PDF in this expression can be obtained from the propagator 
$G_0(\mbox{\boldmath$x$}, \mbox{\boldmath$x$}_0,t)$, which is a joint probability density
for $x_1$ and $x_2$ conditional upon $t$, by dividing by the appropriate marginal density of $x_1$:
\begin{equation}
\label{eq: 4.3}
P_{x_2|t}(x_2,t)=\frac{G_0 \left((0,x_2),(x_0,0),t) \right)}
{\int_{-\infty}^\infty {\rm d}x_2\ G_0 \left((0,x_2),(x_0,0),t) \right)}
\ .
\end{equation}
Using equation (\ref{eq: 2.1}), since $t \ge 0$,  we can write this as
\begin{equation}
\label{eq: 4.4}
P_{x_2|t}(x_2,t) = \frac{ \exp{ \left[-S (\mbox{\boldmath$\xi$},t ) \right] }}{I( x_0,t)}
\end{equation}
where  $\mbox{\boldmath$\xi$} = (-x_0, x_2)$ and 
\begin{equation}
\label{eq: 4.5}
I(x_0,t) \equiv \int_{-\infty}^\infty {\rm d} x_2 \ \exp{\left[ -S(\mbox{\boldmath$\xi$},t ) \right]}
\ .
\end{equation}
Then equation (\ref{eq: 4.2}) becomes
\begin{equation}
\label{eq: 4.6}
P_{x_2}(x_2) \sim 
\frac{x_0}{\sqrt{4\pi D_{11}}} 
\int_{0}^\infty \frac{{\rm d}t } { t^{3/2} }
\left[ \frac{1}{ I \left( \xi_1,t \right)}
 \exp{\left( -S\left(\mbox{\boldmath$\xi$},t \right) -
\frac{\left(x_0-v_1 t \right)^2}{4D_{11}t } \right)} \right]
\ .
\end{equation}
In the steady state the $x_1$ component of the particle flux onto the boundary has magnitude  $J_{\rm a}(x_2) =  \sigma_0 P_{x_2} ( x_2)$.

\subsection{Estimates using the method of Laplace}
\label{sec: 4.2}
For large $|\mbox{\boldmath$\xi$}|$ (i.e. for large $|x_2|) $ both of the integrals in equation 
(\ref{eq: 4.6}) can be approximated by the method of Laplace. This allows us to determine 
the asymptotic form of the PDF of the ordinate of the  absorption point, $P_{x_2} ( x_2)$ 
and, thereby, the magnitude of the absorption flux $J_{\rm a}(x_2)$.

For any given values of $\xi_1$ and $t$ the dominant contribution to $I(x_0,t)$ 
arises from a neighbourhood of the critical value $\xi_2^\ast$, at which  
$S(\mbox{\boldmath$\xi$},t)$ has a minimum with respect to $x_2$ and  
\begin{equation}
\label{eq: 4.7}
I(x_0,t) \sim
\exp{ \left[ -S\left( (\xi_{1},\xi_{2}^\ast ),\ t \right) \right] }  
\int_{-\infty}^\infty  {\rm d}\xi_2 \ \exp \left[- \frac{\left( \xi_{2} - \xi_{2}^\ast \right)^2}{2}
\frac{ \partial^2 S} {\partial \xi_{2}^2 } \left( \left( \xi_{1}, \xi_{2}^\ast \right) ,t \right)   \right]
\ .
\end{equation}
Writing  $\mbox{\boldmath$\xi$}^\ast = \left(\xi_1,\xi_{2}^\ast \right)$, the critical 
value  $\xi_{2}^{\ast} $ satisfies
\begin{equation}
\label{eq: 4.8}
\frac{\partial S}{\partial \xi_{2}} \left( \mbox{\boldmath$\xi$}^\ast,t \right) = 
\frac{{\rm D}_{11} \xi_{2}^\ast - {\rm D}_{12} \xi_{1} - 
\left({\rm D}_{11}  v_{2} - {\rm D}_{12} v_{1} \right) t }{2  \ \rm{det}  \left( \bf {D} \right)  t} =  0 
\end{equation}
so that 
\begin{equation}
\label{eq: 4.9}
\xi_2^\ast = \frac{{\rm D}_{12}}{{\rm D}_{11}} \left( \xi_1 - v_1t \right) + v_2t 
\end{equation}
and
\begin{equation}
\label{eq: 4.10}
\mbox{\boldmath$\xi$}^\ast - \mbox{\boldmath$v$} t = 
\left[ \left( \xi_1 - v_1 t \right), \ \frac{{\rm D}_{12}}{{\rm D}_{11}}\left( \xi_1 - v_1 t \right)  \right]^T
\end{equation}
giving
\begin{equation}
\label{eq: 4.11}
S \left( \mbox{\boldmath$\xi$}^\ast,t  \right) = 
\frac{ \left( \xi_1 - v_{1} t \right)^2}{4{\rm D}_{11} t}
\end{equation}
and
\begin{equation}
\label{eq: 4.12}
\frac{\partial^2 S}{\partial \xi_{2}^2}\left(\mbox{\boldmath$\xi$}^\ast,t \right)  
= \frac{{\rm D}_{11}}{2 \ {\rm det} \left( {\bf D} \right)  t }
\ .
\end{equation}
Substituting these results into  equation (\ref{eq: 4.7})  gives
\begin{equation}
\label{eq: 4.13}
I ( x_0,t ) \sim \exp{ \left[ - \frac{ \left( \xi_1 - v_{1} t \right)^2}{4{\rm D}_{11} t} \right] }
\int_{-\infty}^\infty {\rm d}\xi_{2}  
\exp \left[ - \frac{{\rm D}_{11} \left( \xi_{2} -\xi_{2}^\ast  \right)^2}{4 \ {\rm det}\left( {\bf D} \right) \ t} \right]
\end{equation}
and evaluating the Gaussian integral we have 
\begin{eqnarray}
\label{eq: 4.13a}
I(x_0,t)\sim  \sqrt{ \frac{4 \pi \ {\rm det}\left( {\bf D} \right) \ t}{{\rm D}_{11} }} \ 
\exp{ \left[ - \frac{ \left( \xi_1 - v_{1} t \right)^2}{4{\rm D}_{11} t} \right] }
\ .
\end{eqnarray}
Equation (\ref{eq: 4.6}) now gives
\begin{equation}
\label{eq: 4.14}
P_{x_2}(x_2) \sim 
\frac{x_0}{4 \pi \sqrt{{\rm det}\left( {\bf D} \right)}} 
\int_{0}^\infty \frac{{\rm d}t}{t^{2}} 
\left[ \exp{\left( -S\left(\mbox{\boldmath$\xi$},t \right) -
       \frac{\left(x_0-v_1 t \right)^2}{4{\rm D}_{11}t } 
	     + \frac{\left(\xi_{1}-v_1 t \right)^2}{4{\rm D}_{11}t } \right)}	
		 \right]
\ .		
\end{equation}
Since  $\xi_1 = - x_0$  on the boundary the last two terms in the exponent sum to a constant value giving 
\begin{equation}
\label{eq: 4.15}
P_{x_2}(x_2) \sim \frac{x_0}{4 \pi \sqrt{{\rm det}\left( {\bf D} \right)}}  
\exp  \left( \frac{  x_0 v_1 }{{\rm D}_{11}} \right) \int_{0}^\infty \frac{{\rm d}t}{t^2} 
\exp \left[ -S \left( \mbox{\boldmath$\xi$},t \right) \right] 
\ . 
\end{equation}
The remaining integral may be estimated as in section (\ref{sec: 2.2}) giving
\begin{equation}
\label{eq: 4.16}
\left[ \mbox{\boldmath$J$}_a  \left( x_2 \right) \right]_{1} =
\sigma_0 P_{x_2}(x_2) \sim  \sigma_0 A_{\rm a}( x_2)
     \exp \left[ -\Phi_0 \left( \mbox{\boldmath$x$},\mbox{\boldmath$x$}_0 \right) \right] 
\  
\end{equation}
where $\Phi_0$  is as given by equation (\ref{eq: 2.16}) and
\begin{equation}
\label{eq: 4.17}
A_{\rm a}( x_2) =   
 \left[ 
 \frac{x_0 \exp( x_0 v_1 /{\rm D}_{11})}{2 \sqrt{ \pi {\rm det}( {\bf D})}} 
 \right]
\frac{ \left( \mbox{\boldmath$v$}\cdot {\bf D}^{-1}\mbox{\boldmath$v$} \right)^\frac{1}{4}}
{ \left( \mbox{\boldmath$\xi$}\cdot {\bf D}^{-1}\mbox{\boldmath$\xi$} \right)^\frac{3}{4}}
\ . 
\end{equation}
On the $x_2$ axis far from the source $A_{\rm a}(x_2) \sim |x_2|^{-3/2}$ 
so that $A_{\rm a}$, and the flux onto an absorbing boundary, 
$J_{\rm a}(x_2)$, decays more rapidly than the flux onto a permeable 
boundary, $J_0$, as $|x_2|$ increases. However, surprisingly, the exponent $\Phi_0$
in $J_{\rm a}$ is the same as that in $J_0$.  

\section{Discussion of results}
\label{sec: 5}

The above derivation depends upon using the Laplace method to estimate 
integrals, and it is instructive to assess the accuracy by comparison with Monte Carlo simulations.  The particle paths can be generated using the Euler iterative scheme
\begin{equation}
\label{eq: 5.1}
\mbox{\boldmath$x$}_{n+1} = \mbox{\boldmath$x$}_{n} + \mbox{\boldmath$v$} \delta t 
+ \sqrt{2  \ {\bf D} \ \delta t \ } \mbox{\boldmath$\eta$}_{n}
\end{equation}
with starting value $\mbox{\boldmath$x$}_{0}$,  where  $\delta t$ is a small time 
increment and each $\mbox{\boldmath$\eta$}_{n}$ is a $d$-dimensional vector of 
independent normally distributed random variables with zero mean and unit variance. 

\begin{figure*}[H]
\begin{overpic}[width=1.0\textwidth, grid=false, tics=10]{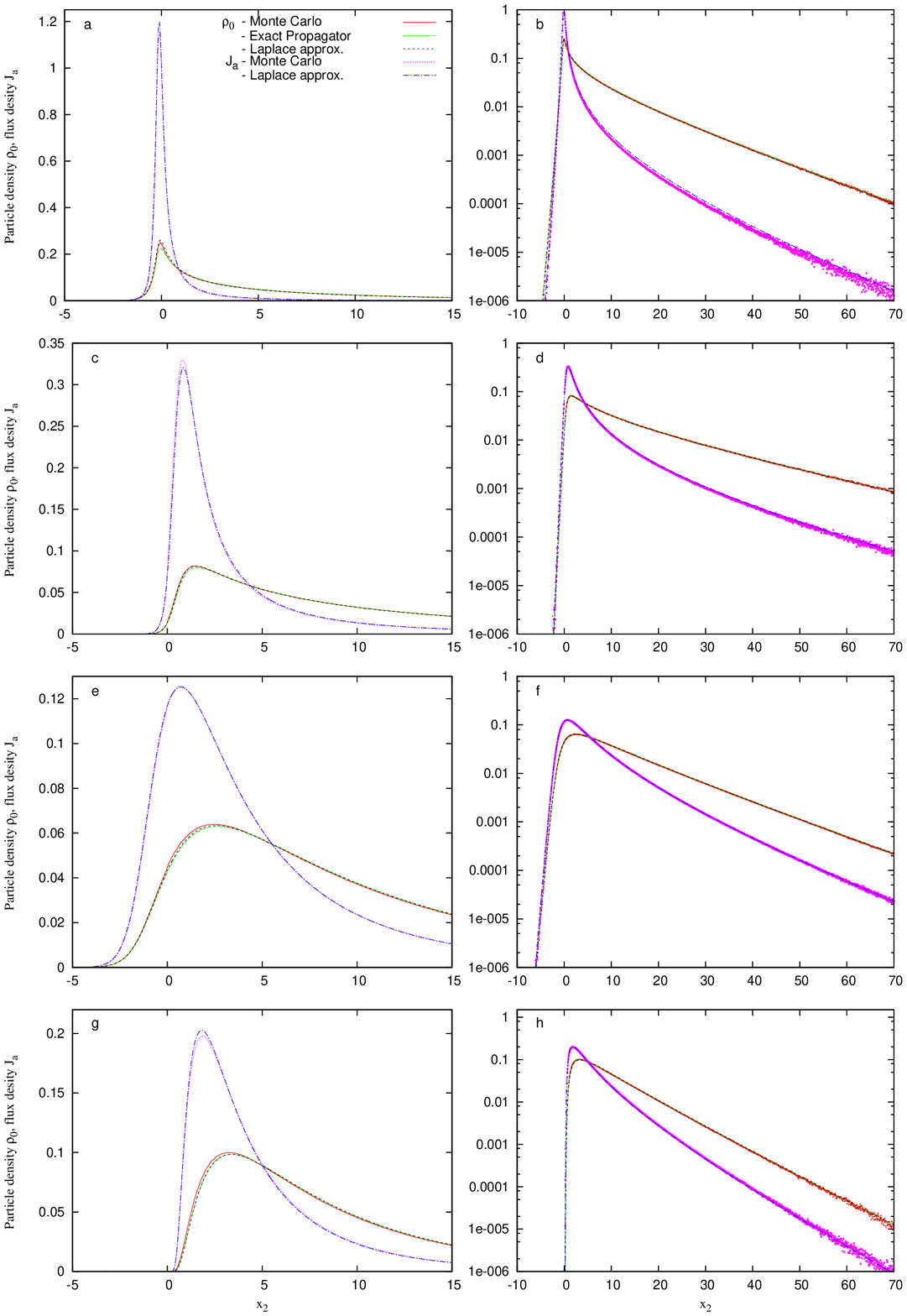}
{\scriptsize
    \put(20,88){$\begin{aligned}
		                      x_0 & =  0.1 \\
          \mbox{\boldmath$v$} & =   \begin{bmatrix} -1 \\ 2 \end{bmatrix}\\
                      {\bf D} & =   \begin{bmatrix} 1.25 & -0.75 \\ -0.75 & 1.50 \end{bmatrix}
		              \end{aligned}$}
    \put(20,68){$\begin{aligned}
		                      x_0 & =  0.25 \\
          \mbox{\boldmath$v$} & =  \begin{bmatrix} -1 \\ 2 \end{bmatrix} \\
                      {\bf D} & =   \begin{bmatrix} 1.25 & -0.75 \\ -0.75 & 1.50 \end{bmatrix}
		              \end{aligned}$}		
    \put(20,42){$\begin{aligned}
		                      x_0 & =  2.5 \\
          \mbox{\boldmath$v$} & =  \begin{bmatrix} -1 \\ 2 \end{bmatrix} \\
                      {\bf D} & =   \begin{bmatrix} 1.25 & -0.75 \\ -0.75 & 1.50 \end{bmatrix}
		              \end{aligned}$}	
    \put(20,16){$\begin{aligned}
		                      x_0 & =  0.25 \\
          \mbox{\boldmath$v$} & =  \begin{bmatrix} -1 \\ 2 \end{bmatrix} \\
                      {\bf D} & =   \begin{bmatrix} 1.00 & 0.00 \\ 0.00 & 0.001 \end{bmatrix}
		              \end{aligned}$}			
}					
\end{overpic}
\caption{\label{fig: 2}(Colour online). Linear and log-linear plots of the magnitudes of the 
steady state particle density for a permeable boundary, $\rho_0(x_2)$ , and the steady-state flux
for an absorbing boundary, $J_a(x_2)$. The plots show normalised densities or fluxes 
at $(0,x_2)$ from a steady source at $(x_0,0)$, for various values of $\mbox{\boldmath$v$}$ 
and {\bf D} and $x_0$.}
\end{figure*}

Figure \ref{fig: 2} shows comparative plots of the magnitudes of the steady state 
absorption flux, ${J_{\rm a}(x_2)}$,  and 
the particle density $\rho_0(x_2)$ for a permeable boundary (the flux $J_0(x_2)$ 
onto a permeable boundary is obtained by multiplying $\rho_0(x_2)$ by an effective drift velocity: 
see equation (\ref{eq: 2.11})). The values of the source location 
$\mbox{\boldmath$x$}_0$, drift velocity $\mbox{\boldmath$v$}$ and diffusion tensor 
${\bf D}$ are given in each of the plots, which  display the same data on linear and 
semi-logarithmic scales side-by-side.  Both $J_{\rm a}(x_2)$ and $\rho_0(x_2)$ are 
displayed as normalised distributions: in the case of $J_{\rm a}(x_2)$ this corresponds to 
the source intensity $\sigma_0$ being equal to unity. The value of $\rho_0(x_2)$ computed 
from (\ref{eq: 2.8}) by numerical integration is also shown.

In plots (a)-(f)  $\mbox{\boldmath$v$} = [ -1,  2 ]^T$ and since $v_2\ne 0$ the distributions 
are skewed. Plots (g) and (h) illustrate the conclusion from section (\ref{sec: 2.3}) that our analysis remains valid even if ${\bf D}$ has rank one.  The curves in the log-linear plots become parallel at large $x_2$ showing that, asymptotically, the exponential decay is the same whether or not there is an absorbing boundary.  

In the case of the non-absorbing boundary, equation (\ref{eq: 2.22}) is expected to serve as a 
precise asymptotic estimate of $\rho_0(x_2)$ as $|x_2|\to \infty$, and the plots in figure (\ref{fig: 2}) 
show that this formula is indeed very accurate. In the case of the absorbing boundary, the 
status of equations (\ref{eq: 4.16}) and (\ref{eq: 4.17}) is more complicated, because there are 
two Laplacian integrations involved in their derivation. These equations are not precise 
asymptotes for $J_{\rm a}(x_2)$ as $|x_2|\to \infty$ because this limit does not play a role in 
the estimate for the normalisation factor of the integral in the denominator of equation 
(\ref{eq: 4.3}), (i.e. equation (\ref{eq: 4.13a})). Equations (\ref{eq: 4.16}), (\ref{eq: 4.17}) are 
only precise asymptotic estimates when we take $x_0\to \infty$ as well as $|x_2|\to \infty$. 
Our numerical results illustrate this: in the case of an absorbing boundary the quality of the 
asymptotic approximation improves as $x_0$ becomes larger.  However the results also show 
that in practice equations (\ref{eq: 4.16}) and (\ref{eq: 4.17}) are accurate even for quite small 
$x_0$ and that they predict the correct scaling of $A_{\rm a}(x_2)$.

In order to obtain $J_{\rm a}(x_2)$ we have used an \lq antiparticle' approach to determine 
the distribution of first contact times for the boundary, and combined this with a calculation 
of the distribution $P_{x_2|t}$ of the coordinate $x_2$ conditional upon the first contact time 
$t$. It is natural to ask whether the \lq antiparticle' picture could lead to a more direct evaluation 
of $J_{\rm a}(x_2)$ in the case where the distance of the source from the boundary, $x_0$, is small. 
In that case most of the particles emitted from the source at $\mbox{\boldmath$x$}_0=(x_0,0)$ 
will be absorbed by the boundary at a point downstream of the source, 
$\mbox{\boldmath$x$}_{\rm d}=\mbox{\boldmath$x$}_0+\mbox{\boldmath$v$}x_0/v_1$. 
One might expect that the flux at $\mbox{\boldmath$x$}$ would then be given by a 
dipole approximation
\begin{eqnarray}
\label{eq: 5.2}
J_{\rm a}(\mbox{\boldmath$x$},\mbox{\boldmath$x$}_0)&\sim &
J_0(\mbox{\boldmath$x$},\mbox{\boldmath$x$}_0)-
J_0(\mbox{\boldmath$x$},\mbox{\boldmath$x$}_{\rm d})
\nonumber \\
&\sim & (\mbox{\boldmath$x$}_0-\mbox{\boldmath$x$}_{\rm d}) 
\cdot \mbox{\boldmath$\nabla$}_{\small{\mbox{\boldmath$x$}_0}}  
J_0 (\mbox{\boldmath$x$},\mbox{\boldmath$x$}_0) 
\end{eqnarray}
however we were not able to produce a valid estimate of $J_{\rm a}(x_2)$ using this type of 
dipolar approximation.

In conclusion, we have investigated the flux $J_{\rm a}(x_2)$ onto an absorbing boundary, 
as a function of the coordinate $x_2$ measuring the distance from the source. We find 
that $J_{\rm a}(x_2)\sim A_{\rm a}(x_2)\exp[-\Phi_0(x_2)]$, where the exponent $\Phi_0(x_2)$ 
grows linearly as $x_2\to \infty$ and is the same as the exponent for the flux onto a permeable 
boundary. The function $A_{\rm a}(x_2)$ was shown to have an algebraic decay: 
$A_{\rm a}(x_2)\sim |x_2|^{-3/2}$ for the absorbing boundary, compared to 
$A_0(x_2)\sim |x_2|^{-1/2}$ for a permeable boundary. These results will be directly 
relevant to extending recent studies \cite{Wil+14} of the geometry of constellations of 
points sampling fractal measures.

This research was supported by the NORDITA program {\sl Dynamics of Particles in Flows}. 

{}

\end{document}